\documentclass[runningheads,a4paper]{llncs}

\usepackage{amssymb}
\usepackage{graphicx}

\usepackage{url}
\urldef{\mailsa}\path{ mathcci@yahoo.com}
\urldef{\mailsb}\path {honghyejong@yahoo.com}
\urldef{\mailsc}\path|erika.siebert-cole, peter.strasser, lncs}@springer.com|

\begin{document}

\mainmatter  

\title{Formal Design and Verification of N-M Switching Control System}

\titlerunning{Formal Design and Verification of N-M Switching Control System}

%
%
\author{Changil Choe$^1$, Hyejong Hong$^2$, and Kukhwan Kim$^3$}
\authorrunning{Changil Choe, Hyejong Hong, and Kukhwan Kim}

\institute{$^{1, 3}$Faculty of Mathematics, Kim Il Sung University, D.P.R.K\\
\mailsa\\
$^2$Central Information Agency of Science and Technology, D.P.R.K\\
\mailsb
}

%
%

\toctitle{Lecture Notes in Computer Science}
\tocauthor{Authors' Instructions}
\maketitle

\begin{abstract}
Production factories in which stable voltage is critical, e.g., electro-plating factory, require constantly stable voltage to minimize loss by adjusting incoming voltage in real time, even if low-quality electricity is supplied from outside. To solve such problem often being raised from the factories located in the area with unstable electricity supply, we designed N-M switching control system and verified its correctness using LTL model checking technique.
\end{abstract}

\section{Introduction}
Production under unstable electricity condition may cause serious loss of expense, such as large amount of rejects, in factories that include processes requiring highly stable electricity. For example, professional precious metal plating companies require to keep stable voltage at all times they work. In case that such factories or companies are situated in the unstable power supply area, voltage stabilization is raised more importantly. 

N-M switching control system was motivated to meet the requirement raised from a plating factory for eating utensil set [3]. The quality of electricity being supplied to the factory was not good, that is, some days low or high voltage electricity was supplied. There was a manual voltage regulator switching between M levels in factory. When incoming voltage is too low or high, however, it was impossible to supply necessary voltage to the whole workplaces of the factory, even if operator switched up to maximum or minimum level. This is not a problem that would be solved just by installing larger capacity regulator, because bandwidth of input voltage of each regulator is limited.

From the standpoint of profit of the factory, it was better to produce only in workplaces which can be supplied with normal voltage by adjusting and distributing incoming voltage in real time, rather than all workplaces were exposed to the production failure of acceptable goods. But, there was no voltage regulator meeting such special requirement (See Section 2 for more details), though numerous works were devoted to design and implement various types of stable voltage suppliers, e.g., [1, 2]. To solve the problem we designed N-M switching control system, shortly N-M system in this paper, by combining a PC with the manual voltage regulator of the factory.
N-M system is a real-time voltage normalization and distribution system that divides whole workplaces of factory into N sections, adjusts voltage height by switching between $M$ levels, and supplies normalized voltage to sections according to the given priority. To the best of our knowledge, N-M system is an original work, and thus design verification is necessarily needed for the successful implementation of system at low cost. 

In this paper, we describe the working mechanism of N-M system and present a method to verify correctness of its design using LTL model checking technique. LTL (Linear-time Temporal Logic) is a kind of temporal logic having strong expressive power to specify time-dependant properties of real-time systems and LTL-based model checking technique is now widely used in verification of real-time systems [4, 5, 6, 7, 8]. We don’t concentrate on describing the details of N-M system and the whole specifications of its requirements, rather focus on showing our method to verify N-M system for its time-dependant requirements using LTL model checking technique. 

\section{Working Mechanism and Design Requirements}

In this section, we describe working mechanism, implementation method and time-dependant requirements of N-M system.

Whole workplaces of factory are divided into $N$ sections $W_1, W_2 , \ldots, W_N$ by considering relative independence of work. Power supply priority is assigned to each section according to the importance or processing order of products. For example, we may give highest priority to silver-plating workplace. For the convenience of description, we assume that $W_i$ has higher priority than $W_j$ if $i<j$. Voltage is adjusted at $M$ levels $L_1, L_2 , \ldots, L_M$. There are three states for each level, that is, low voltage state $l$, normal voltage state $n$ and high voltage state $h$. This standard is set by considering technical requirements of production.

 We briefly describe working mechanism of N-M system below. 
System starts control in level $L_\alpha$ where $\alpha=\lceil M/2 \rceil$  and does one of the following three behaviors. 
\begin{itemize}
\item Increase voltage by switching level into $L_{\alpha+1}$, if the incoming voltage is low.

\item  Supply electricity to section $W_1$, if the incoming voltage is normal. 

\item Drop voltage by switching level into $L_{\alpha-1}$, if the incoming voltage is high.
\end{itemize}

Let us now assume that system is in level $L_m$ and current supplying sections are $W_1, W_2 ,\ldots, W_n$.

\begin{itemize}
\item Suspend electricity supply to $W_n$, if the incoming voltage is low.

\item  Supply electricity to section $W_{n+1}$, if the incoming voltage is normal. 

\item  Drop voltage by switching level into $L_{m-1}$, if the incoming voltage is high.
\end{itemize}

For the practical design and implementation, more items than described above must be considered. The purpose of the paper is to show verification method of N-M system, and thus we don’t consider some details of the system. 

Control of N-M system is realized using the values of $N+M+3$ bit string corresponding to the sections $W_1, W_2 , \ldots, W_N$, voltage adjustment level $L_1, L_2 , \ldots,$ $L_M$ and voltage states $l, n, h$ in each level. For example, in case $N=3$ and $M=2$, bit value string 111 10 010 denotes that normal voltage is supplied to all sections by increasing voltage adjustment level to the maximum. 

The number of possible bit value string is $2^{N+M+3}$ for $N+M+3$ bit string, but some bit value string does not occur in control. In the above case, bit strings 010 01 010 and 010 11 001 do not occur. This is because system does not supply electricity to $W_{i+1}$ unless $W_i$ is supplied with electricity and voltage adjustment can not be in different level at the same time. Exact number of bit value strings occurring in control is $(N+1) \times (4 \times M)$. This is not small number and it may fail to implement correct control system if we don't verify design.

We only consider 8 requirements of N-M system for the purpose of the paper, though there are many other requirements to be verified. \\

$D_1$: System decreases work section by one, if the voltage state is low in maximum level.

$D_2$: System suspends electricity supply to all sections, if the voltage state is high in minimum level. 

$D_3$: System keeps current supplying sections and levels up by one, if voltage state is low and leveling up is possible. 

$D_4$: System keeps current supplying sections and levels down by one, if voltage state is high and leveling down is possible.

$D_5$: System increases work section by one, if voltage state is normal in current level.

$D_6$: System keeps current supply, if all sections are supplied with electricity and voltage state is normal in current level. 

$D_7$: System does not supply electricity to $W_{i+1}$ unless $W_i$ is supplied with electricity.

$D_8$: It is possible to supply electricity to all sections. 

\section{Verification of Design using LTL Model Checking}

In this section, we present our method to verify N-M switching control system for its requirements using LTL model checking technique. For this, we construct transition system model of N-M system and write LTL specifications of its requirements. Then we check satisfaction relation between model and specifications using LTL model checking tool NuSMV [9-11]. NuSMV (New Symbolic Model Verifier) was designed as an reliable verification of industrial sized designs and an research tool for formal verification techniques. NuSMV supports the analysis of specifications expressed in LTL and other temporal logic CTL. 

\begin{figure}
\centering
\includegraphics[height=2.3cm]{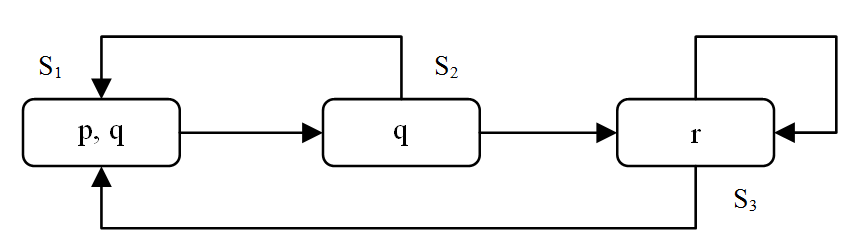}
\caption{A Transition System}
\label{fig1}
\end{figure}

A transition system $\mathcal{M}$ is a triple $\mathcal{M}=(S, T, L)$ consisting of a finite set $S$ of states, a transition relation $T\subseteq S\times S$, a labelling function $L:S\rightarrow 2^{AP}$ which assigns the set of atomic propositions to each state $s \in S$. $AP$ is the set of observable atoms of the system. Fig.~\ref{fig1} shows an example of transition system.

For LTL modeling of N-M system, we use the following atomic propositions. 
\begin{itemize}
\item $W_i (i=1, \ldots,  N)$: Section $W_i$ is supplied with electricity. 

\item  $L_j (j=1, \ldots,  M)$: Voltage adjustment level is $L_j$. 

\item  $l$: Voltage is low in current level.

\item $n$: Voltage is normal in current level. 	

\item $h$: Voltage is high in current level.
\end{itemize}

\begin{figure}
\centering
\includegraphics[height=7.3cm]{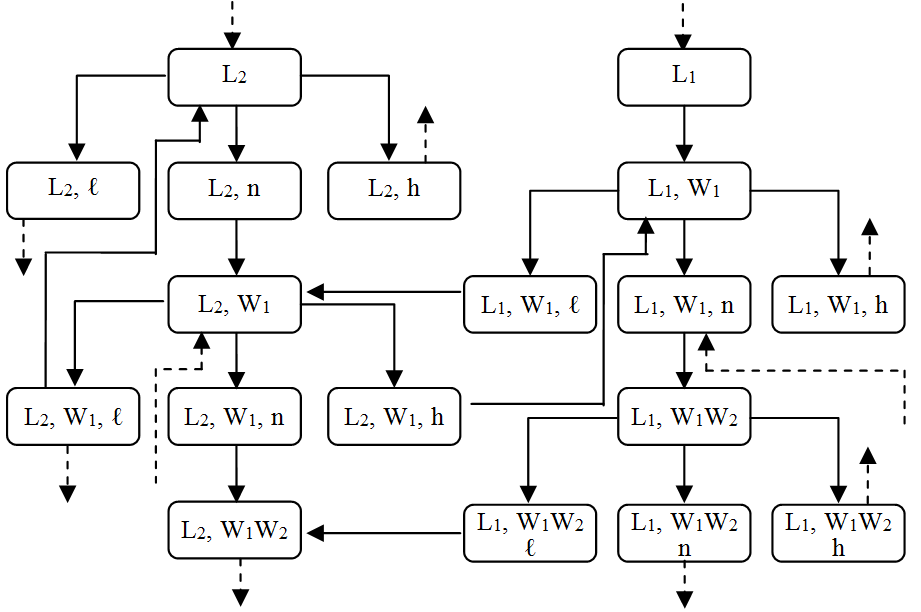}
\caption{Transition System Model of N-M Switching Control System}
\label{fig2}
\end{figure}

As we mentioned in Section 2, total number of states of N-M system is $(N+1) \times (4 \times M)$. It is difficult to draw complete transition system model of N-M system in a page. We only show a part of model in Fig.~\ref{fig2}. 

A LTL formula $\phi$ is built up from a finite set of atomic propositions, the propositional operators $\neg, \land, \lor, \rightarrow$  and temporal modal operators $X, F, G, U, W, R$. Among the temporal operators, $X, F$ and $G$ are used in this paper. LTL formulas are estimated on the path of transition system. Let $s_0 \rightarrow s_1 \rightarrow s_2 \rightarrow\ldots$ be a path of a transition system $\mathcal{M}$. 

\begin{itemize}
\item $X\phi$ means that $\phi$ has to hold at $s_1$ on the path.  

\item $F\phi$ means that $\phi$ eventually has to hold at a state $s_i (i \geq 1)$, somewhere on the path. 

\item  $G\phi$ means that $\phi$ has to hold at all states $s_i (i \geq 0)$ on the path. 
\end{itemize}

Let $s_0$ be the initial state of a transition system $\mathcal{M}$ and $\phi$ be a LTL formula. It is called that $\phi$ is satisfied by $\mathcal{M}$, denoted by $\mathcal{M}, s_0\models \phi$, if $\phi$ holds on every path of $\mathcal{M}$, which starts from $s_0$. Requirements of N-M system can be specified with LTL operator $G, X$ and $F$ as follows. (Note that $D1, D3, D4, D5$ and $D7$ are formula schemata)\\

$D_1: G(L_1\land l \land W_1\land \ldots \land W_i \rightarrow X(W_1\land \ldots \land W_{i-1}))  $ $   $ $   $ $  $ $  $ $  $ $ $ $   $ $   $ $  $ $  $ $  $ $ i=1,\ldots, N$ \\

$D_2: G(L_M \land h \rightarrow X(\neg W_1 \land \ldots  \land \neg W_N))$ \\

$D_3: G(L_j \land l \land W_1 \land \ldots \land W_i \rightarrow X(L_{j+1} \land W_1 \land \ldots \land W_i))$  $ $   $ $   $ $  $ $  $ $  $ $ $ i=1,\ldots, N$ 

$ $ $ $ $ $ $ $ $ $ $ $  and $j=1, \ldots, M-1$   \\

$D_4: G(L_j \land h \land W_1 \land \ldots \land W_i \rightarrow X(L_{j-1} \land W_1 \land \ldots \land W_i))$  $ $   $ $   $ $  $ $  $ $  $ $   $i=1,\ldots,N$
 
$ $ $ $ $ $ $ $ $ $ $ $ and  $j=2, \ldots, M$ \\ 

$D_5: G(n \land W_1 \land \ldots \land W_i \rightarrow X(W_1 \land \ldots \land W_{i+1}))    $ $   $ $ $ $   $ $   $ $  $ $  $ $  $ $ $ $   $ $   $ $  $ $  $ $  $ $ i=1,\ldots,N-1$  \\ 

$D_6: G(n \land W_1 \land \ldots  \land W_N \rightarrow X(W_1 \land \ldots  \land W_N))$\\

$D_7: G\neg(\neg W_i \land W_j)$  $ $   $ $ $ $   $ $   $ $  $ $  $ $  $ $ $ $   $ $   $ $  $ $  $ $  $ $  $ $   $ $ $ $   $ $   $ $  $ $  $ $  $ $ $ $   $ $   $ $  $ $  $ $  $ $  $ $   $ $ $ $   $ $   $ $  $ $  $ $  $ $ $ $   $ $   $ $  $ $  $ $  $ $  $ $   $ $ $ $   $ $   $ $  $ $  $ $  $ $ $ $   $ $   $ $  $ $  $ $  $ $   $ $   $ $ $ $   $ $   $ $  $ $  $ $  $ $ $ $   $ $   $ $  $ $  $ $  $ $  $ $   $ $ $ $   $ $   $ $  $ $  $ $  $ $ $ $   $ $   $ $      $1 \leq i < j \leq N$\\

$D_8$ cannot be specified as a LTL formula directly. The LTL specification of the negation of $D_8$ is as follows. \\

$D'_8: \neg F(W_1 \land \ldots \land W_N)$ \\

Therefore, if $D'_8$ is not satisfied by a transition system, then $D_8$ is satisfied by it and vice versa. \\

Using temporal logic model checker NuSMV, we checked the satisfaction relation between transition system model and LTL specifications of N-M system. Through several executions of NuSMV and debugging, we could construct the transition system model of N-M system satisfying real-time requirements including $D_1, \ldots, D_8$. Using this model, we designed and implemented correct N-M switching control system meeting the requirement of factory.

\section{Conclusion}
Temporal logic model checking is very useful technique to design and implement real-time systems like N-M switching control system. We believe that N-M switching control system and its verification method presented in the paper can be used in other cases of designing and verifying control systems.

\end{document}